\documentclass{PoS}

\title{Towards the NNPDF3.0 parton set for the second LHC run}

\ShortTitle{NNPDF3.0}

\author{\speaker{M.~Ubiali} \thanks{on behalf of the NNPDF collaboration}\\
        DAMTP, Centre for Mathematical Studies, Wilberforce Rd, CB3 0WA\\
       \& Department of Physics, Cavendish Laboratory, J.J. Thomson Avenue, CB3 0HE\\ 
       University of Cambridge, Cambridge, UK\\
        E-mail: \email{ubiali@hep.phy.cam.ac.uk}}

\abstract{The full exploitation of the increasingly precise LHC measurements 
is essential in order to reduce the uncertainty of theoretical predictions at hadron colliders.
The NNPDF2.3 fit was the first PDF determination including the effect of the early LHC data. Here the new NNPDF3.0 PDF set 
is announced and its main features are presented. The novel NNPDF analysis is based on an improved fitting methodology, 
statistically validated by closure tests. Over a thousand new data points are included, both the recent HERA II measurements 
and a wide set of new LHC data. In this contribution details on the experimental data are given and their impact on PDF uncertainties is displayed.}

\FullConference{XXII. International Workshop on Deep-Inelastic Scattering and Related Subjects,\\
		28 April - 2 May 2014\\
		Warsaw, Poland}

\begin{document}

\section{Introduction}
Parton Distribution Functions (PDFs) play a crucial role in the LHC physics program. 
The PDF error is the dominant source of theoretical uncertainty for several key processes at the LHC, 
and therefore it limits the accuracy of both SM analyses, such as the characterisation of the Higgs boson, 
and BSM searches, particularly those involving heavy particles in the final state. 
The excellent performance of the LHC machine asks for a completely new level of precision in PDF analyses, 
especially in view of the high--energy run. 

The NNPDF collaboration has developed a methodology to obtain a reliable determination of PDFs and a robust estimate 
of their error, based on Monte Carlo techniques and on the use of Neural Networks.
In the latest NNPDF analyses, the NNPDF2.3~\cite{nnpdf23} and the NNPDF2.3QED~\cite{nnpdf23qed}, 
several LHC measurements have been included for the first time in a fit. 
Their impact was found to be significant, and particularly crucial to determine the photon PDF. 
The leading order NNPDF2.3QED fit is the new default PDF set in Pythia 8, 
used in the recent Monash tune~\cite{Skands:2014pea}.
 
The NNPDF3.0 analysis represents a major upgrade. First of all, the fit is based on a more efficient code 
completely rewritten in {\tt c++}. Most importantly, over a thousand new data points are analysed, 
both from HERA II and from the LHC.
The significant increase in the number of experimental data is accompanied by a refined fitting 
methodology both based on and validated by a statistical closure test. 
Such test has been proposed for several years~\cite{Lyons:2008zz} as a rigorous way for 
PDF fitters to directly evaluate the efficacy of their fitting methodology and quantify the flexibility 
of their PDF parametrisation by studying a
fit to an idealised data set produced according to a known
underlying physical law. For a detailed discussion about closure tests we refer to~\cite{ClosureTest}.
Finally, a number of upgrades in the theory are
implemented, such as the inclusion of the electroweak corrections in the computation of the observables for which 
the latter are significant, a detailed study of the known NNLO corrections to the jet inclusive cross section and finally
the switch from the FONLL-A to the FONLL-B scheme~\cite{Forte:2010ta} at NLO in order to improve the theoretical description of the charm structure function.

Details on the new experimental data are found in the next section, followed by some preliminary results. 
For all further details on the NNPDF3.0 analysis we refer to the upcoming paper~\cite{30}. 

%%% 
\section{New experimental data}
\begin{figure} [h]
\begin{center}
\includegraphics[width=.8\textwidth]{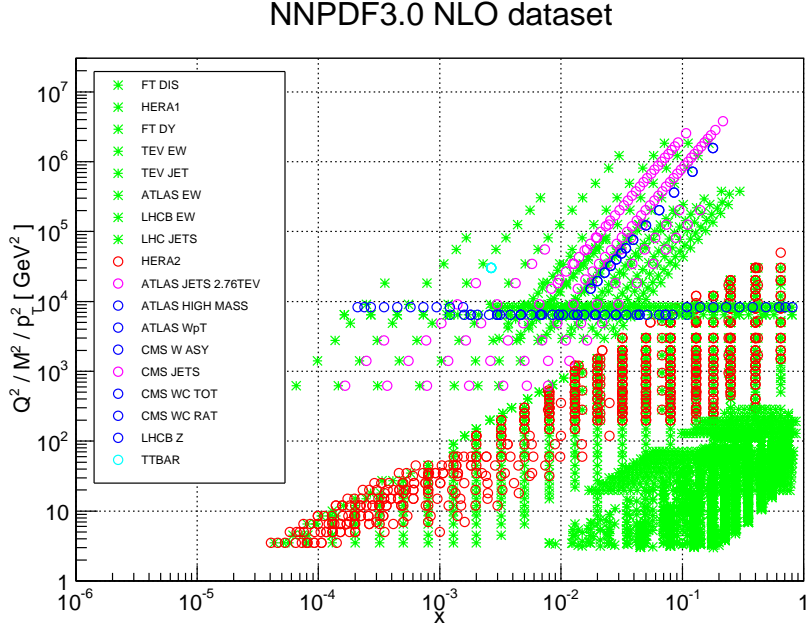} 
\end{center}
\caption{Kinematic coverage in $x$ and $Q^2$ of the experimental data included in the NLO NNPDF3.0 analysis. 
In green are the data points already included in the NNPDF2.3 analysis. In red the new HERA II data included 
in NNPDF3.0, in blue the LHC data on vector boson production, in purple the LHC jet data and in light blue the 
point corresponding to the total top pair production cross section.} 
\label{kin}
\end{figure}

The kinematical coverage of the new HERA and LHC data included in the NLO NNPDF3.0 analysis is displayed in Fig.~\ref{kin}. 
The NNLO dataset is slightly different, due to different kinematic cuts.
The recent HERA-II measurements of inclusive cross sections %~\cite{HERAIIa,Collaboration:2010xc,Collaboration:2010ry,Abramowicz:2012bx}
and the combined determination of the charm production cross section %~\cite{Abramowicz:1900rp}, 
come with an improved statistical and systematic precision, 
thus providing a stronger handle on quarks at medium- and large-$x$ and on the gluon at small-$x$.
Once the full combined HERA-I and II data will be publicly available, 
the full combined set can be consistently included in future analyses.

In our previous PDF determination, despite the limited statistics of the early measurements, the impact of the LHC data 
turned out to be moderate but non negligible. In the past few months, several new measurements have been made 
available by the LHC experiments, 
based on a larger statistic sample and with smaller systematic uncertainties. 
They are expected to provide a unique handle on PDFs. On top of the vector boson production data from 
ATLAS, LHCb and CMS that were already present in NNPDF2.3 we include:
the ATLAS high-mass Drell-Yan production data%~\cite{Aad:2013iua}
, based on an integrated luminosity of 4.9 fb$^{-1}$, 
which give important information on the large-$x$ quark-antiquark separation;
the transverse momentum distribution of $W$ bosons %~\cite{Aad:2011fp} 
measured by ATLAS, which we expect to provide a complementary constraint on the gluon in the
medium-$x$ region; 
the CMS charged $W$ muon asymmetry based on the full statistics (5 fb$^{-1}$) of the 7 TeV run%~\cite{Chatrchyan:2013mza}
, which displays substantially reduced statistical and systematic uncertainties
with respect to the previous measurement in the electron channel;
the 7 TeV double differential Drell-Yan cross sections%~\cite{Chatrchyan:2013mza}
, which cover a large range of 
lepton pair invariant mass and provide an important handle on quark-flavor separation in a wide range of $x$. 
Finally we analyse the $W$ production in association with charm quarks measured by CMS, %~\cite{Chatrchyan:2013uja}, 
and the 940 pb$^{-1}$ forward $Z\to e^+e^-$ production data taken at LHCb in 2011. %~\cite{Aaij:2012mda}.
The former allows to directly probe the strangeness, which is the worst known of all light quark PDFs and which at the moment is 
mostly determined from the old NuTeV and CHORUS fixed target data. The second provides a precious handle on the poorly known
very small and very large $x$ regions.\\
%
% JETS %
As far as jets are concerned, we add the 5 fb$^{-1}$ 7 TeV inclusive cross section dataset from CMS, %~\cite{Chatrchyan:2012bja},
and the 2.76 TeV dataset from ATLAS, %~\cite{Aad:2013lpa}
supplemented by its correlation with the 7 TeV data, %~\cite{Aad:2011fc} 
which were already included in our previous analysis. 
Finally, we include the four measurements of the total top pair production cross section at 7 and 8 TeV provided by the ATLAS and CMS collaborations.
This observable is known at NNLO~\cite{Czakon:2013goa} and it is free of any non-perturbative ambiguity. 
The tight constraint provided by this data on the gluon at large-$x$ was displayed in previous works~\cite{Czakon:2013tha,Beneke:2012wb} 
and confirmed by our analysis.

% Theory settings
We conclude this section with a short description of the improved theory settings of our analysis. 
The remarkably high accuracy of the experimental measurements requires theoretical predictions
at the percent level of precision. 
At such level of accuracy, nor the NLO EW contributions to the Drell-Yan pair production
or the corrections beyond NLO to the jet inclusive cross section can be neglected.
In the large invariant mass region, at which the high mass ATLAS and CMS measurements are taken,
the EW corrections can reach $\mathcal{O}(10\%)$~\cite{Boughezal:2013cwa}.
We implement them via the computation of electroweak $C$-factors with {\tt FEWZ3.1}~\cite{Li:2012wna}.
The code enables us to separate the guage--invariant QED subset of the corrections from the full EW result. 
We isolate and keep into account only 
the weak portion of the EW corrections, that does not require the fit of an initial photon PDF. 
A further effort has been invested in setting up a prescription that enables us to consistently 
include jet data in our NNLO fit without spoiling its theoretical accuracy.
Indeed, the full NNLO contribution has been computed only for the gluon--gluon channel~\cite{Ridder:2013mf}, but is not
yet available for the other channels involving initial quarks. We have an approximate NNLO prediction 
available, which keeps into account threshold effects~\cite{deFlorian:2013qia}. As a result of a
detailed comparison between the full and approximated corrections,
we keep only a subset of the jet data, which are selected by requiring that the exact $C$-factors in the 
$gg$ channel do not differ from the $C$-factors computed with the threshold approximation by more than 15\%.

\section{Results and conclusions}
\begin{figure} [t]
\begin{center}
\includegraphics[width=.45\textwidth]{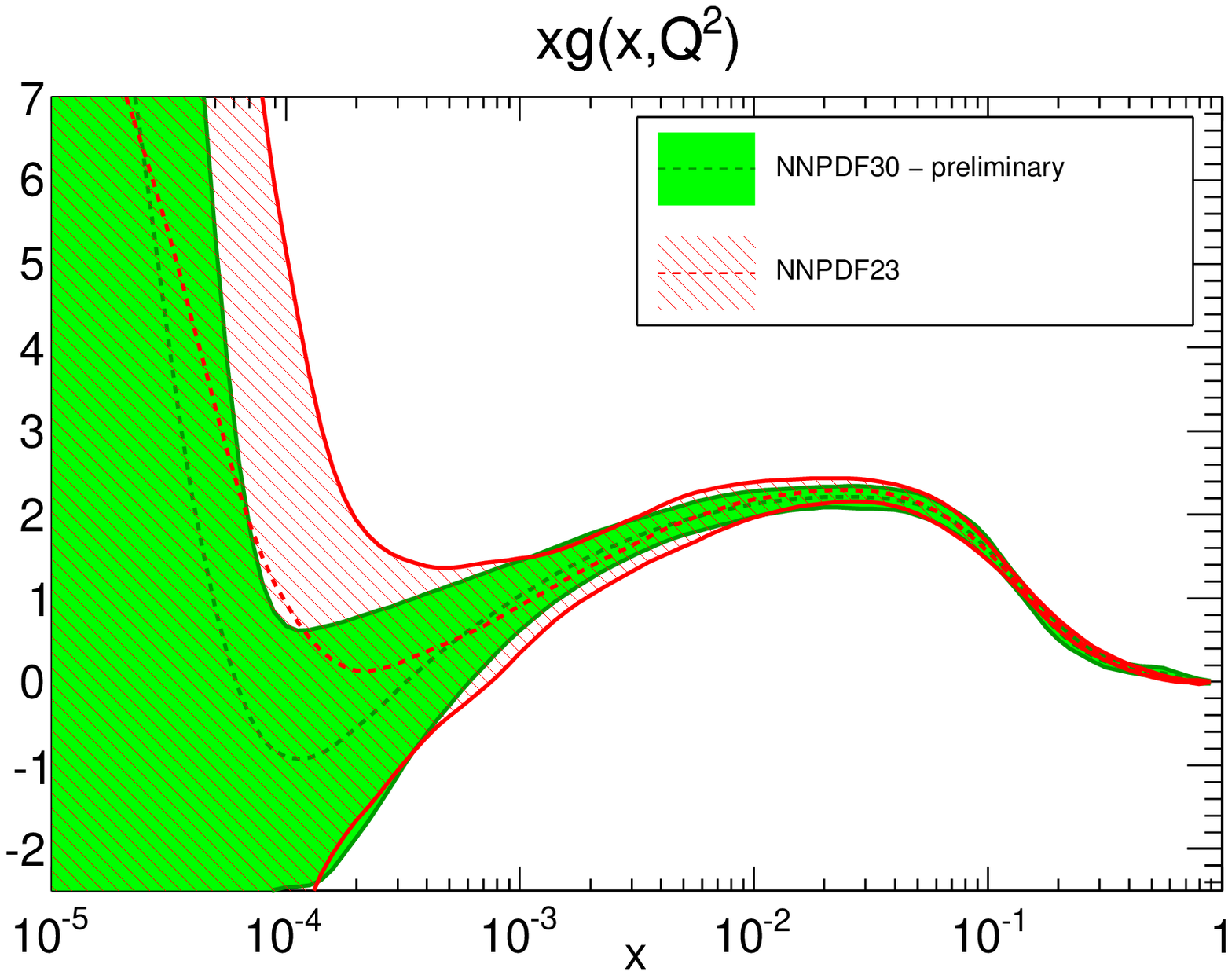} 
\includegraphics[width=.45\textwidth]{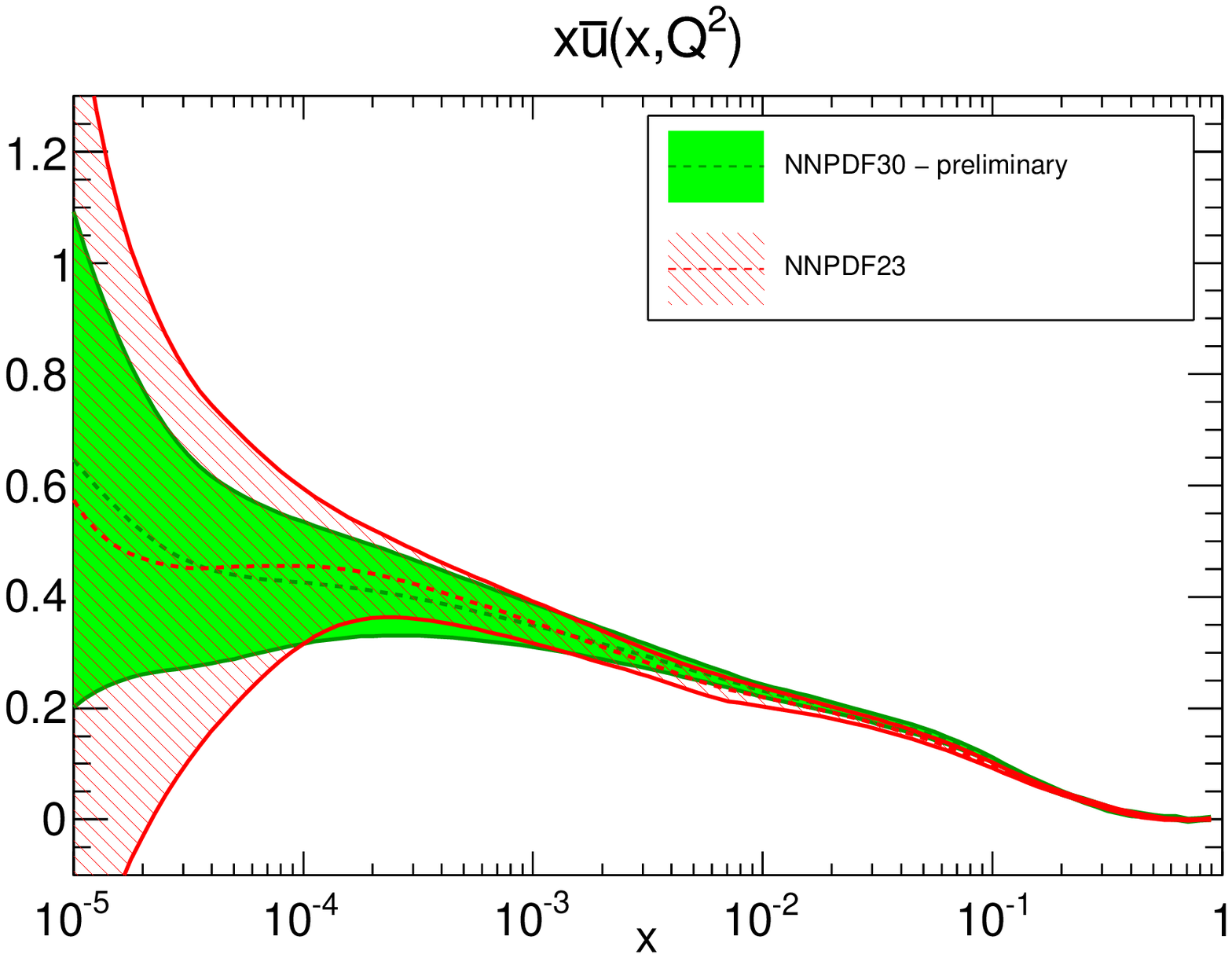} 
\end{center}
\caption{Preliminary NNLO NNPDF3.0 gluon (left plot) and $\bar{u}$ (right plot) at $Q^2 = 2 \,{\rm GeV}^2$ compared 
to their NNPDF2.3 counterparts computed using $N_{\rm rep} = 100$ replicas
from both sets. $\alpha_s(M_Z)=0.118$. All error bands shown correspond to one sigma interval.} 
\label{parton}
\end{figure}
In Fig.~\ref{parton} we show a preliminary comparison between the NNPDF2.3 and the NNPDF3.0 NNLO fits. 
The effect of the inclusion of the new HERA and LHC data is sizeable. The recent jet data from ATLAS and CMS, 
the total top pair production cross section measurements and the new data on the charm structure function provide a powerful handle 
and yield a significant reduction in the gluon uncertainty for a wide range of $x$. 
At the same time, the inclusion of several new precise measurements of vector boson production, 
particularly the high and low mass Drell-Yan data and the precise muon charge asymmetry data from CMS 
help in constraining light quark and antiquark distributions, as it is observed in the case of the anti-up 
distribution displayed in Fig.~\ref{parton}. \\
Clearly the inclusion of the new data is not the only factor that must be take into account when looking at the results.  
The improved fitting methodology and the wider dataset are the two key ingredients for the new precise 
determination of PDFs. Their effects will be explained in more details and disentangled in~\cite{30}, 
in which also a number of crucial phenomenological implications of our analysis, concerning the determination of the proton strangeness, 
the shift in central values and the improved accuracy for several key processes at the LHC are discussed. 
The implications for Higgs production at the LHC will be illustrated and an up-to-date comparison with other PDF determinations 
will be provided.

\end{document}